\documentclass[prb,letterpaper,twocolumn]{revtex4-1}
\usepackage{bbm,color,amsmath,amssymb,graphicx}
\usepackage[pdftex]{hyperref}
\usepackage{braket}

\begin{document}
\title{Dynamics of two coupled semiconductor spin qubits in a noisy environment}
\author{S.\ Das Sarma}
\author{Robert E.\ Throckmorton}
\author{Yang-Le Wu}
\affiliation{Condensed Matter Theory Center and Joint Quantum Institute, Department of Physics, University of Maryland, College Park, Maryland 20742-4111, USA}
\date{\today}
\begin{abstract}
We theoretically consider the temporal dynamics of two coupled spin qubits (e.g., semiconductor quantum
dots) driven by the inter-qubit spin-spin coupling.  The presence of environmental noise (e.g., charge
traps, nuclear spins, random magnetic impurities) is accounted for by including random magnetic field and
random inter-qubit coupling terms in the Hamiltonian.  Both Heisenberg coupling and Ising coupling between
the spin qubits are considered, corresponding respectively to exchange and capacitive gates as appropriate
for single spin and singlet-triplet semiconductor qubit systems, respectively.  Both exchange (Heisenberg)
and capacitive (Ising) coupling situations can be solved numerically exactly even in the presence
of noise, leading to the key findings that (i) the steady-state return probability to the initial state
remains close to unity in the presence of strong noise for many, but not all, starting spin configurations,
and (ii) the return probability as a function of time is oscillatory with a characteristic noise-controlled
decay toward the steady-state value.  We also provide results for the magnetization dynamics of the coupled
two-qubit system.  Our predicted dynamics can be directly tested in the already existing semiconductor spin
qubit setups providing insight into their coherent interaction dynamics.  Retention of the initial state spin
memory even in the presence of strong environmental noise has important implications for quantum computation
using spin qubits.
\end{abstract}
\maketitle

\section{Introduction}
Universal quantum computation depends on controllable (and highly accurate) one- and two-qubit gate operations
using a suitable physical system of two-level quantum objects (i.e., the qubits).  Among the many different
physical qubit systems being studied worldwide (e.g., ion traps, neutral atoms, superconducting qubits, spin
qubits), both involving atomic and solid-state architectures, semiconductor nanostructure-based spin qubits
are perhaps unique in terms of their easy controllability and fast gate operations, and most importantly, in
terms of their potential for scalability because of existing semiconductor-based microelectronics technology
\cite{R1}.  The subject of semiconductor spin qubits, which for our purposes in the current work are localized
electron spins in semiconductor nanostructures acting as quantum two-level systems, is currently one of the most
active research areas in quantum information processing using solid-state materials.  In fact, both Si- and
GaAs-based spin qubit architectures are being extensively studied in many laboratories all over the world with
several breakthrough experiments being reported during the last five or so years\cite{R2,R3,R4,R5,R6,R7,R8,R9,R10,R11}.
In particular, experimental two-qubit systems using localized electron spins have recently been reported for
both GaAs and Si nanostructures\cite{R12,R13,R14}, although the fidelity for such two-qubit gate operations is
still rather low due to environmental noise and weak inter-qubit coupling.

The current theoretical work is specifically on a two-qubit system of localized electron spins in semiconductor
nanostructures (e.g., Si or GaAs quantum dots, P or other donor states in Si, Si MOS-based qubits, Si-Ge heterostructure-based
qubits) with the goal being to understand the quantum dynamics of the two-qubit system in the presence
of both inter-qubit coupling and environmental noise.  The environmental noise could arise from a number of physical
mechanisms, but the most important ones for semiconductor spin qubits are known to be Overhauser noise due to
the background nuclear spins and charge noise due to traps and defects in the system.  The noise is represented
in the theory by a random magnetic field acting on single qubits (i.e., each spin) and a random fluctuation in
the two-spin coupling (i.e., the two-qubit interaction).  The goal is to understand the behavior of coupled-qubit
spin dynamics in the presence of both inter-qubit interaction and environmental noise.

Our theory is primarily motivated by the fact that there are very few laboratory demonstrations of controlled two-qubit
coupling in semiconductor spin quantum computing experiments.  Even the few that exist\cite{R12,R13,R14} typically
manifest very poor fidelity and sometimes even poor reproducibility\cite{R15}.  This is in sharp contrast with the
competing quantum computing architectures involving superconducting and ion trap qubits, where multi-qubit coherent
quantum control operations are now routinely achieved in many laboratories with high ($>95\%$) fidelity and reproducibility.
Since eventual large-scale quantum information processing would require efficient and controllable one- and two-qubit
gates with very high ($>99.9\%$) fidelity, the paucity of two-qubit gate operations in semiconductor spin quantum
computing systems is particularly worrisome, and is now universally considered to be the key roadblock for future
progress in the field of semiconductor-based quantum computation.

Given the above context of the critical absence of two-qubit coupling demonstration experiments in semiconductor spin
quantum computing architectures, it is perhaps imperative to take a step back and consider situations where the
two-qubit entanglement dynamics can be directly explored without the complication of gate control operations in
order to understand the interplay between qubit coupling and qubit noise.  In the current work, therefore, we study
theoretically the coupled qubit dynamics of two localized electron spin qubits in semiconductor nanostructures
in the presence of finite qubit coupling and finite noise.  The theoretical problem allows for an almost exact
analytical solution, making the results tractable and leading to specific predictions which are directly testable
in the currently existing experimental spin qubit systems.  We believe that the experimental demonstration of the
theoretical two-qubit coupling dynamics of the current work would be easier than the full two-qubit gate control
experiments, providing useful insight into the interplay between interaction and noise in the dynamics of coupled
spin qubits.  This is because a two-qubit gate control experiment would not only require multiple measurements of
return probabilities, one for each member of a suitable orthogonal basis, e.g., $\ket{\uparrow\uparrow}$, $\ket{\uparrow\downarrow}$,
$\ket{\downarrow\uparrow}$, and $\ket{\downarrow\downarrow}$, but also requires a means of error correction, which
can result in complicated pulse sequences.  On the other hand, demonstrating the results of our work simply requires
a single measurement of the return probability, and does not require any error correction scheme---one simply lets
the system evolve as it will under the influence of both an intentionally applied gate voltage and the noise in
the system.  Another key difference between performing a quantum computing gate operation and our proposal is that
performing a gate operation requires precise control over the timing and duration of pulses, whereas our proposal
only involves a simple evolution in time with constant ``always on'' gate voltages, which is convenient for an
experimentalist.

Experimentally, semiconductor spin qubits form essentially a $2\times 2$ matrix in laboratory implementations with
both Si and GaAs being used as the materials platform and there being two distinct types of qubit architectures
and coupling, namely, the so-called exchange\cite{LossPRA1998,DiVincenzoNature2000,HuPRA2000,ScarolaPRA2005} and
capacitive\cite{CalderonVargasPRB2015,SrinivasaPRB2015} two-qubit gates, in operation.  The two-qubit exchange
gate involves a direct Heisenberg coupling between two localized electron spins whereas the two-qubit capacitive
gate involves a dipolar capacitive Ising-type coupling between two quantum dot based singlet-triplet qubits\cite{LevyPRL2002,PettaScience2005}.
We refer to these couplings as Heisenberg (i.e., exchange) and Ising (i.e., capacitive) throughout this paper, and
we consider them on an equal footing although their experimental implementations involve different physics and
architectures.

In addition to the interqubit coupling, the qubits are affected by environmental noise arising from several
different mechanisms\cite{HuPRL,WitzelPRL,CywinskiPRL,BarnesPRL,CywinskiPRB} that we parametrize in our theory
by random local magnetic fields acting on both qubits (in addition, of course, to any known applied field for
qubit control).  We also include a noise term in the qubit coupling itself so that the theory has both one-
and two-qubit noise.  {There has been previous work considering magnetic noise\cite{CoishPRB2005,KlauserPRB2006},
as well as some limited work on charge noise\cite{KlauserPRB2006}; however, the latter does not explore the effects
of charge noise on the detailed dynamics of the system, and thus we undertake such an investigation here.  In
particular, the goal of our work is to directly motivate experimental investigations of coupled two-qubit dynamics
in recently realized experimental systems\cite{R2,R3,R4,R5,R6,R7,R8}.  The main point of the paper is the interplay
between the qubit coupling and the qubit noise (both one- and two-qubit) on the temporal dynamics of the coupled
qubits, which can be solved essentially analytically, leading to precise experimentally observable consequences.
These experiments to verify our predicted coupled qubit dynamics can be performed right now, both for Heisenberg
and Ising coupling gates and both in Si and GaAs spin qubit platforms---the necessary laboratory systems for carrying
out the qubit dynamics experiments already exist.  The experimental situation to be used here is what is commonly
referred to as the ``always on'' configuration, in which the inter-qubit coupling is not turned off at all after
it has been turned on.

In Sec.\ II we review the Heisenberg and Ising models and their diagonalization.  In Secs.\ III
and IV, respectively, we provide our (analytical and numerical) results for the Heisenberg and Ising qubit couplings.
We conclude in Sec.\ V with a summary and outlook.

\section{Models}
Below, we review the theoretical models for Heisenberg (Sec.\ II A) and Ising (Sec.\ II B) interqubit coupling on an
equal footing using the same notation as much as possible, as well as the eigenvalues and eigenvectors for the Heisenberg
case\cite{CoishPRB2005,KlauserPRB2006}.  We review the corresponding results for the Ising case\cite{WangNPJQE} in the
Appendix.  Throughout this paper, we will be working in units in which $\hbar=1$.

\subsection{Heisenberg}
The Heisenberg Hamiltonian describing two coupled spin qubits in the presence of qubit noise is given by the following
disordered Heisenberg Hamiltonian model:
\begin{equation}
H=J\vec{S}_1\cdot\vec{S}_2+h_1 S_{1z}+h_2 S_{2z},\label{Eq:HeisenbergH}
\end{equation}
where the sites, $1$ and $2$, denote the two localized electron spins (i.e., qubits) with an interqubit coupling strength
$J$ that could have a small random component $\delta J$ in it denoting the noise-induced fluctuation in the interqubit
coupling.  The quantities, $h_1$ and $h_2$, are the environmental noise-induced random magnetic fields at the two qubits.
In principle, we could add a constant non-random externally applied uniform magnetic field $h_0$ to Eq.\ \eqref{Eq:HeisenbergH}
by redefining $h_1\to h_0+h_1$ and $h_2\to h_0+h_2$, but, as will become clear below, this will have no effect on the
dynamics that we describe in this work, and is therefore left out (the situation is different for the Ising model, in
which such an external magnetic field will affect the dynamics, making the Ising problem, to be discussed below in Sec.\
II B, richer and more complex than the Heisenberg case).  These noise-induced local magnetic fields, arising from background
nuclei and charge noise (and possibly from other unknown sources), are taken to be Gaussian random variables (with no loss
of generality---other choices for the random noise distribution do not change any of our conclusions).  We assume the noise
to be static since it is slow in reality (and its exact time dependence is unknown), but we discuss later in the paper the
considerations for dynamic noise.

We note that the total spin $S_z=S_{1z}+S_{2z}$ for the two-spin system is conserved and focus on the $S_z=0$ subspace,
which consists of only two states $\ket{\uparrow\downarrow}$ and $\ket{\downarrow\uparrow}$, where the arrows indicate
the $z$ component of the spin (up or down) for each qubit.  The reason for restricting ourselves to $S_z=0$ is that the
other two possibilities, $S_z=\pm 1$ (for a two-spin Heisenberg system, there are only three possible $S_z$ subspaces,
defined by $S_z=0,\pm 1$), are trivial since the system is now stuck in a single eigenstate with no dynamics whatsoever.

{In the $S_z=0$ basis, we can rewrite the two-qubit Hamiltonian as\cite{CoishPRB2005,KlauserPRB2006}}
\begin{equation}
H=\tfrac{1}{2}J\sigma_z+\tfrac{1}{2}\delta h\sigma_x-\tfrac{1}{4}J,\label{Eq:HeisenbergHSz0}
\end{equation}
where $\sigma_x$ and $\sigma_z$ are the usual $2\times 2$ Pauli spin matrices in the $S_z$ basis and $\delta h$ is the
noise-induced random field difference between qubits $1$ and $2$:
\begin{equation}
\delta h=h_1-h_2.\label{Eq:MagFieldDiff}
\end{equation}
We note that $J$ by itself may include a random term $\delta J$ as well.  Note that any constant uniform applied magnetic
field drops out of $\delta h$ leaving only the noise-induced random field difference between the two qubits (this simplification
does not exist for the Ising case as we will see later).  Here, we use the convention that the upper component of the wave
function corresponds to $\ket{\uparrow\downarrow}$.

{The eigenvalue problem for the Hamiltonian defined in Eq.\ \eqref{Eq:HeisenbergHSz0} is exactly solvable\cite{CoishPRB2005,KlauserPRB2006},
giving the following energy eigenvalues $E_{\pm}$ and eigenstates $\psi_{\pm}$:}
\begin{equation}
E_{\pm}=-\tfrac{1}{4}J\pm\tfrac{1}{2}\sqrt{J^2+(\delta h)^2}\label{Eq:HeisenbergHEvs}
\end{equation}
and
\begin{equation}
\psi_{\pm}=\begin{bmatrix}
\frac{1}{\sqrt{2}}\sqrt{1\pm\frac{\delta h}{\sqrt{J^2+(\delta h)^2}}} \\
\pm\frac{1}{\sqrt{2}}\sqrt{1\mp\frac{\delta h}{\sqrt{J^2+(\delta h)^2}}}
\end{bmatrix}.\label{Eq:HeisenbergHEss}
\end{equation}
We use these equations to obtain the results presented in Sec.\ III.

\subsection{Ising}
The Ising Hamiltonian describing two coupled spin qubits (e.g., singlet-triplet quantum dot qubits) in the presence of
qubit noise is given by
\begin{equation}\label{Eq:IsingH}
H=\varepsilon J_1 J_2\sigma_1^z\sigma_2^z+J_1\sigma_1^z+J_2\sigma_2^z+h_1\sigma_1^x+h_2\sigma_2^x,
\end{equation}
where we have expressed the coupled Hamiltonian in the so-called ``singlet-triplet basis,'' with $\sigma_1^z=\pm 1$ for
the singlet/triplet state of the qubit.  In Eq.\ \eqref{Eq:IsingH}, $J_i$ ($i=1,2$) is the individual exchange coupling
within the $i$th qubit (usually a two-quantum dot structure) producing the singlet-triplet single qubit (which necessitates
an intra-qubit exchange coupling in order to create singlet and triplet levels---this exchange coupling is not the interqubit
coupling here, it is merely a parameter defining the qubit itself) whereas the interqubit coupling is given by $\varepsilon J_1 J_2$
with $\varepsilon$ denoting the interqubit coupling strength.  The noise-induced random magnetic fields $h_i$ ($i=1,2$) act
on each qubit by coupling through $\sigma_i^x$ in the singlet-triplet basis.

Comparing Eq.\ \eqref{Eq:IsingH} with Eq.\ \eqref{Eq:HeisenbergHSz0}, we note significant differences between the Ising and
Heisenberg models, with the Ising model being much more complex.  In particular, the Ising Hamiltonian, Eq.\ \eqref{Eq:IsingH},
is a $4\times 4$ matrix in the singlet-triplet basis whereas the Heisenberg Hamiltonian, Eq.\ \eqref{Eq:HeisenbergH} or
\eqref{Eq:HeisenbergHSz0}, is simply an (effective) $2\times 2$ matrix.  One consequence of the richer structure of the
Ising coupling (and the associated singlet-triplet semiconductor spin qubits) is that the total (effective) $S_z$ is not conserved,
in contrast with the Heisenberg coupling, and therefore the Ising Hamiltonian cannot be expressed in a block diagonal form
[i.e., $\sigma_i^x$ and $\sigma_i^z$ both appear in Eq.\ \eqref{Eq:IsingH}] in contrast to the Heisenberg Hamiltonian,
which is readily expressed in a block diagonal form in the conserved $S_z$ basis.  The more complicated nature of the
capacitive Ising coupling for the singlet-triplet qubits is also reflected in the larger ($3$) number of parameters ($J_1$,
$J_2$, $\varepsilon$) necessary to define the qubit Hamiltonian compared with the Heisenberg Hamiltonian, for which the
simple exchange coupling between the two spins is described by just one parameter, the Heisenberg exchange coupling $J$.

It turns out, however, that in spite of the complicated nature of the Ising Hamiltonian, one can still diagonalize it
analytically, leading to an explicit construction of the time evolution operator controlling the qubit dynamics.
Unfortunately, this analytical solution is built upon the roots of a quartic equation\cite{WangNPJQE} and does not
have an easily transparent (or manageable) form for actual calculations.  We therefore review this analytical
solution in the Appendix of the current paper.

We note that in general one can think of the random on-site magnetic field disorder to be arising from the nuclear
Overhauser noise and the noise in the exchange coupling itself to be arising from the charge noise.  This qualitative
distinction between Overhauser and charge noise applies to both Heisenberg and Ising coupling situations.

\section{Heisenberg coupling}
We first present our detailed results for the Heisenberg Hamiltonian, Eqs.\ \eqref{Eq:HeisenbergH} and \eqref{Eq:HeisenbergHSz0}.
We will consider two starting states---the ``classical'' state, $\ket{\uparrow\downarrow}$, and the singlet state,
$\ket{S}=\frac{1}{\sqrt{2}}(\ket{\uparrow\downarrow}-\ket{\downarrow\uparrow})$.  Results for the qubit dynamics depend
strongly on the initial state.

\subsection{``Classical'' starting state}
We start by giving the return probability and the magnetization at each site, with ``magnetization'' here simply
meaning the expectation value of the spin at the given site.  Taking our initial state to be $\ket{\psi(0)}=\ket{\uparrow\downarrow}$,
we find that the state at time $t$, $\ket{\psi(t)}$, is
\begin{eqnarray}
\ket{\psi(t)}&=&\frac{1}{\sqrt{2}}\sqrt{1+\frac{\delta h}{\sqrt{J^2+(\delta h)^2}}}e^{-iE_+t}\ket{+} \cr
&+&\frac{1}{\sqrt{2}}\sqrt{1-\frac{\delta h}{\sqrt{J^2+(\delta h)^2}}}e^{-iE_-t}\ket{-},
\end{eqnarray}
where $\ket{+}$ and $\ket{-}$ are the positive- and negative-energy eigenstates within the $S_z=0$ basis, respectively.
{We then recover the expression for the return probability found in Ref.\ \onlinecite{CoishPRB2005}:}
\begin{eqnarray}
P_{\uparrow\downarrow}(t)&=&\left |\braket{\psi(t)|\psi(0)}\right |^2 \cr
&=&1-\frac{J^2}{J^2+(\delta h)^2}\sin^2\left [\tfrac{1}{2}\sqrt{J^2+(\delta h)^2}t\right ]\label{Eq:RetProb_UD}
\end{eqnarray}
and the $z$ component of the magnetization of spin $1$ is
\begin{eqnarray}
\braket{S_{1z}(t)}&=&\braket{\psi(t)|S_{1z}|\psi(t)}=\tfrac{1}{2}\left\{\frac{(\delta h)^2}{J^2+(\delta h)^2}\right. \cr
&+&\left.\frac{J^2}{J^2+(\delta h)^2}\cos\left [\sqrt{J^2+(\delta h)^2}t\right ]\right\}. \nonumber \\ \label{Eq:Mag1_UD}
\end{eqnarray}
The magnetization of spin $2$ is just $\braket{S_{2z}(t)}=-\braket{S_{1z}(t)}$.  We provide plots of the return probability and
magnetization for several values of $\delta h$ in Fig.\ \ref{Fig:SingRealUD}.
\begin{figure}[ht]
\includegraphics[width=\columnwidth]{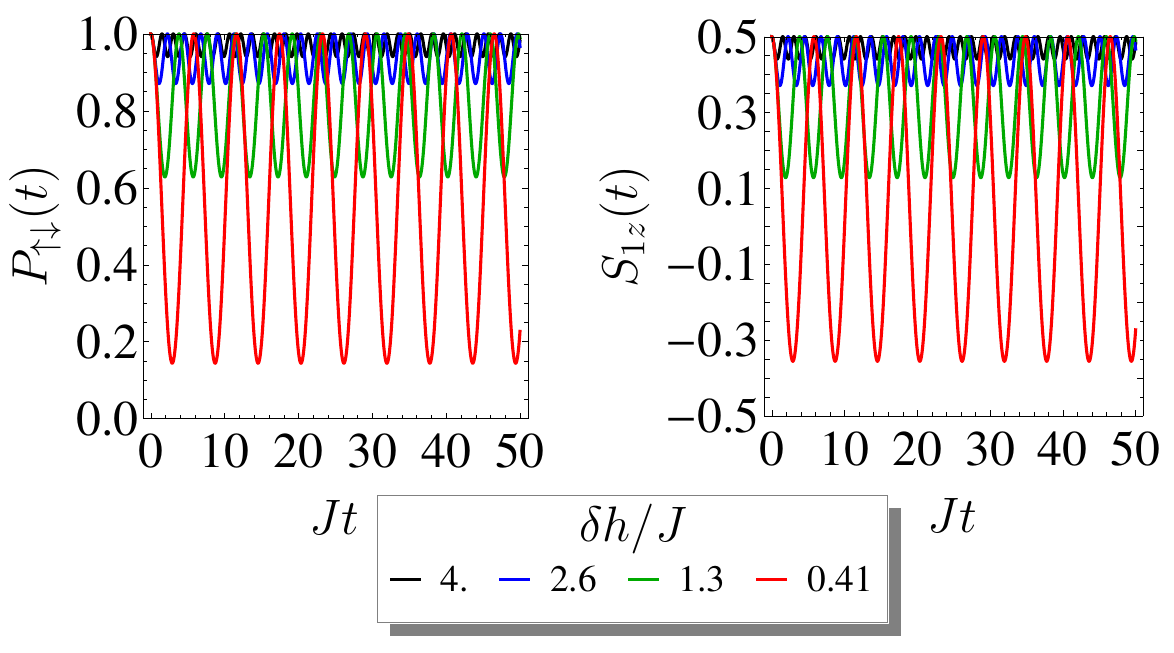}
\caption{Plot of the return probability, $P_{\uparrow\downarrow}(t)$ (left), and of the magnetization of spin $1$,
$\braket{S_{1z}(t)}$ (right), as a function of time for several values of $\delta h$ with the initial state $\ket{\psi(0)}=\ket{\uparrow\downarrow}$.}
\label{Fig:SingRealUD}
\end{figure}

We now determine the disorder-averaged return probability.  We will consider both cases with only magnetic disorder,
i.e., disorder in the magnetic fields $h_1$ and $h_2$, and cases with exchange disorder as well, i.e., disorder in the
exchange coupling $J$.  We will assume that the magnetic fields both follow a Gaussian distribution with zero mean and
standard deviation $\sigma_h$; we will refer to the latter as the ``strength'' of the magnetic disorder.  Similarly, we
will assume that the exchange coupling follows a truncated Gaussian distribution, restricted to non-negative values,
with mean $J_0$ and standard deviation, or strength, $\sigma_J$.   Since the return probability and magnetization only
depend on $\delta h$, we may simplify this calculation by noting that $\delta h$ also follows a Gaussian distribution, with
zero mean and standard deviation $\sigma_h\sqrt{2}$.  Results for the cases in which there is a large magnetic field gradient
present (i.e., the average field gradient $\delta h_0 \gg \sigma_h$) and in which the exchange coupling is very large
(i.e., $J\gg\delta h_0$ and $\sigma_h$), both with $\sigma_J=0$, have been previously found in Ref.\ \onlinecite{KlauserPRB2006}.

The disorder average of some quantity $A$ is given by
\begin{equation}
\left [A\right ]_{\alpha}=\int_{0}^{\infty}\,\int_{-\infty}^\infty dJ\,d(\delta h)\,f_{\delta h}(\delta h)f_J(J)A,\label{Eq:DisorderAvg}
\end{equation}
where the probability density $f_{\delta h}(\delta h)$ is given by
\begin{equation}
f_{\delta h}(\delta h)=\frac{1}{2\sigma_h\sqrt{\pi}}e^{-(\delta h)^2/4\sigma_h^2}\label{Eq:MagDisorderDist}
\end{equation}
and $f_J(J)$ is given by
\begin{equation}
f_J(J)=\frac{1}{\sigma_J\sqrt{2\pi}}\frac{2}{1+\mbox{erf}\left (\frac{J_0}{\sigma_J\sqrt{2}}\right )}e^{-(J-J_0)^2/2\sigma_J^2}.\label{Eq:ExchDisorderDist}
\end{equation}
This integral must be determined numerically.  Using Eqs.\ \eqref{Eq:RetProb_UD} and \eqref{Eq:Mag1_UD}, we evaluate this
integral for the return probability and the magnetization of spin $1$, respectively, for several values of $\sigma_h$ and
$\sigma_J$ and present the results in Figs.\ \ref{Fig:RPSJUD} and \ref{Fig:Mag1SJUD}.  We see that the return probability
and the magnetization oscillate around, and decay to, a steady-state value.  We may argue that the period of the oscillations
is of order $1/J_0$ by noting that the probability distributions for the magnetic and exchange disorder, Eqs.\ \eqref{Eq:MagDisorderDist}
and \eqref{Eq:ExchDisorderDist}, are peaked at $0$ and $J_0$, respectively, and that the oscillation frequency of the oscillatory
terms in the single-realization results given in Eqs.\ \eqref{Eq:RetProb_UD} and \eqref{Eq:Mag1_UD} is $\sqrt{J^2+(\delta h)^2}$.
Therefore, contributions to the disorder average of frequency $J_0$ will dominate the disorder average.  We also note that
the decay rate is set by the disorder strengths $\sigma_h$ and $\sigma_J$, increasing as we increase them.  The magnetic disorder
strength, in addition, sets the steady-state value; we note that this steady-state value increases as we increase $\sigma_h$.
The exchange disorder strength also has a small effect, but it is very small compared to the effect of magnetic disorder.
\begin{figure}[ht]
\includegraphics[width=\columnwidth]{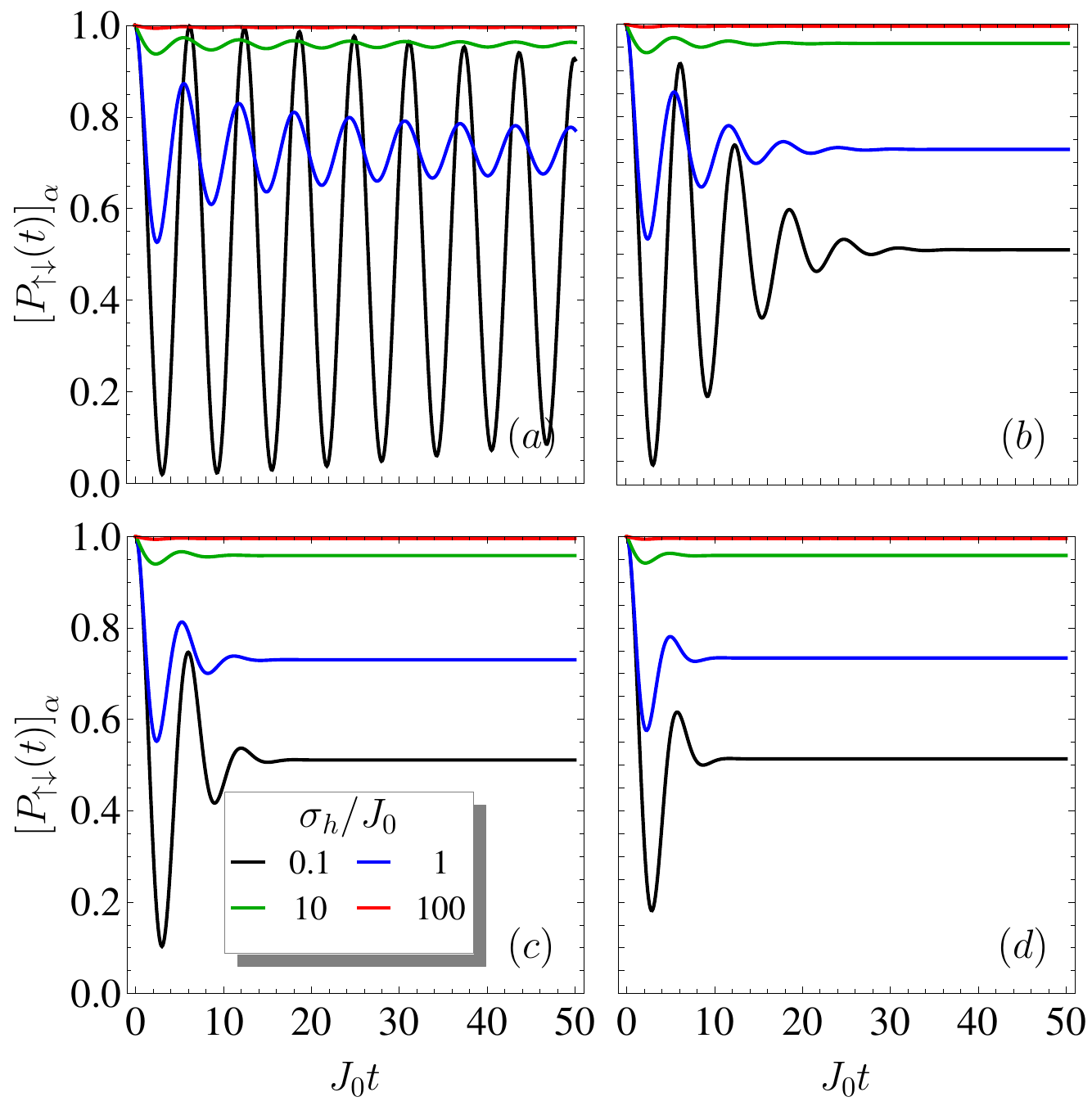}
\caption{Plot of the disorder-averaged return probability, $\left [P_{\uparrow\downarrow}(t)\right ]_{\alpha}$, as a function
of time for different values of $\sigma_h$, as indicated in the plot, with the initial state $\ket{\psi(0)}=\ket{\uparrow\downarrow}$, for
(a) $\sigma_J=0$, (b) $\sigma_J=0.1J_0$, (c) $\sigma_J=0.2J_0$, (d) $\sigma_J=0.3J_0$.}
\label{Fig:RPSJUD}
\end{figure}
\begin{figure}[ht]
\includegraphics[width=\columnwidth]{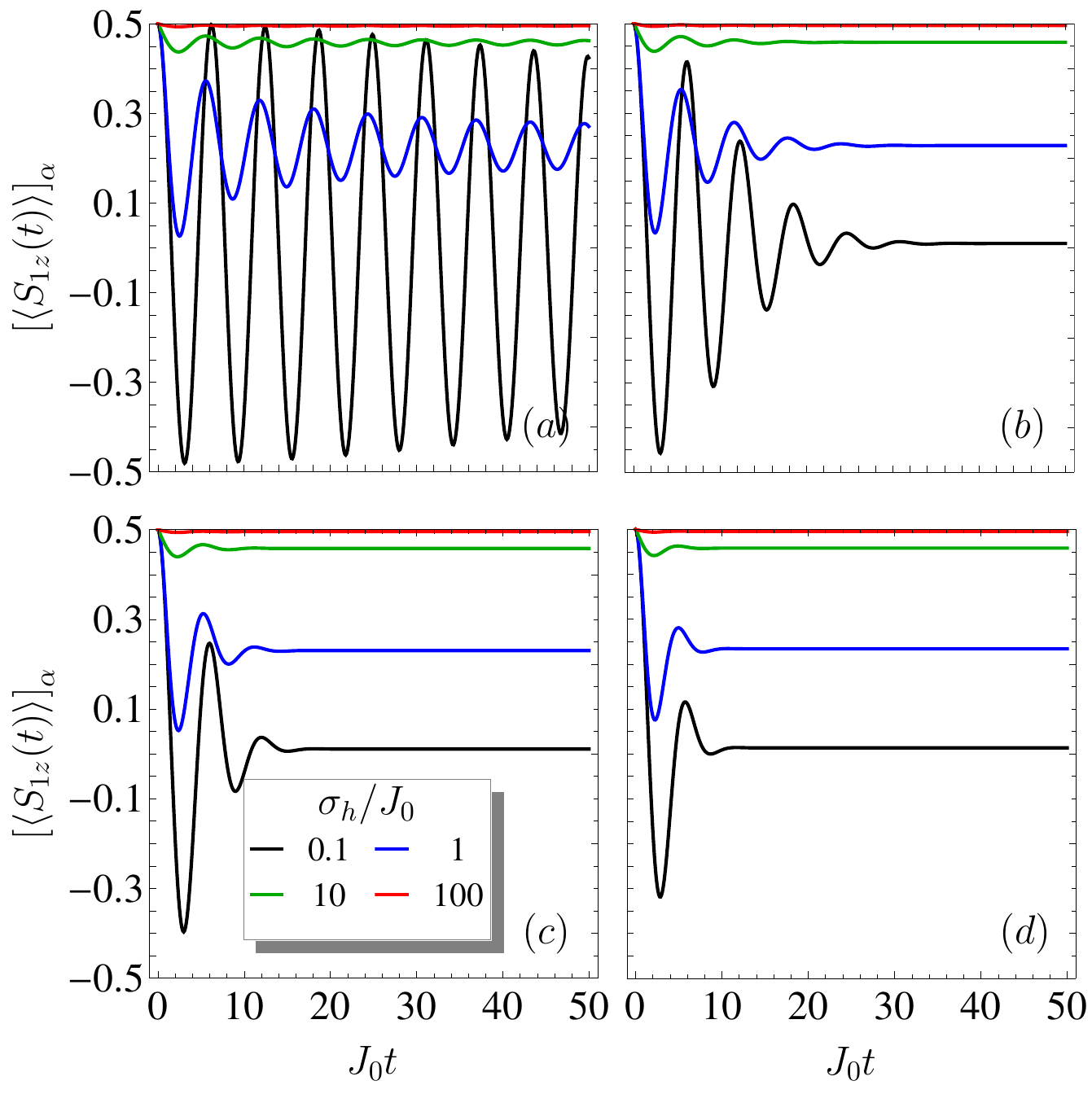}
\caption{Plot of the disorder-averaged magnetization of spin $1$, $\left [\braket{S_{1z}(t)}\right ]_{\alpha}$, as a function
of time for different values of $\sigma_h$, as indicated in the plot, with the initial state $\ket{\psi(0)}=\ket{\uparrow\downarrow}$, for
(a) $\sigma_J=0$, (b) $\sigma_J=0.1J_0$, (c) $\sigma_J=0.2J_0$, (d) $\sigma_J=0.3J_0$.}
\label{Fig:Mag1SJUD}
\end{figure}
We determine the steady-state values as a function of $\sigma_h$ and for $\sigma_J=0$ and plot them in Fig.\ \ref{Fig:LT_noSigJ}.
We have also done calculations for $\sigma_J\neq 0$, and have verified that the results do not differ significantly from the
$\sigma_J=0$ results.  We also confirm that, for sufficiently large values of $\sigma_h$, there is an increase in the steady-state
values of the return probability and the magnetization of spin $1$, as noted above.  This indicates that strong magnetic disorder
helps to preserve the state of the two-qubit system, with the crossover into this behavior happening for $\sigma_h$ on the order
of $J_0$.
Of course, in the situation in which the disorder is much larger than the exchange coupling ($\sigma_h\gg J_0$), this result is
trivial since the interqubit interaction is simply unable to flip the initial spins, but as emphasized elsewhere for multiqubit
systems\cite{BarnesPRB2016}, the return probability is finite even in the situation when the disorder is not much larger than the
coupling, which is neither an intuitive nor an obvious result, and thus should be confirmed experimentally.
\begin{figure}[ht]
\includegraphics[width=0.49\columnwidth]{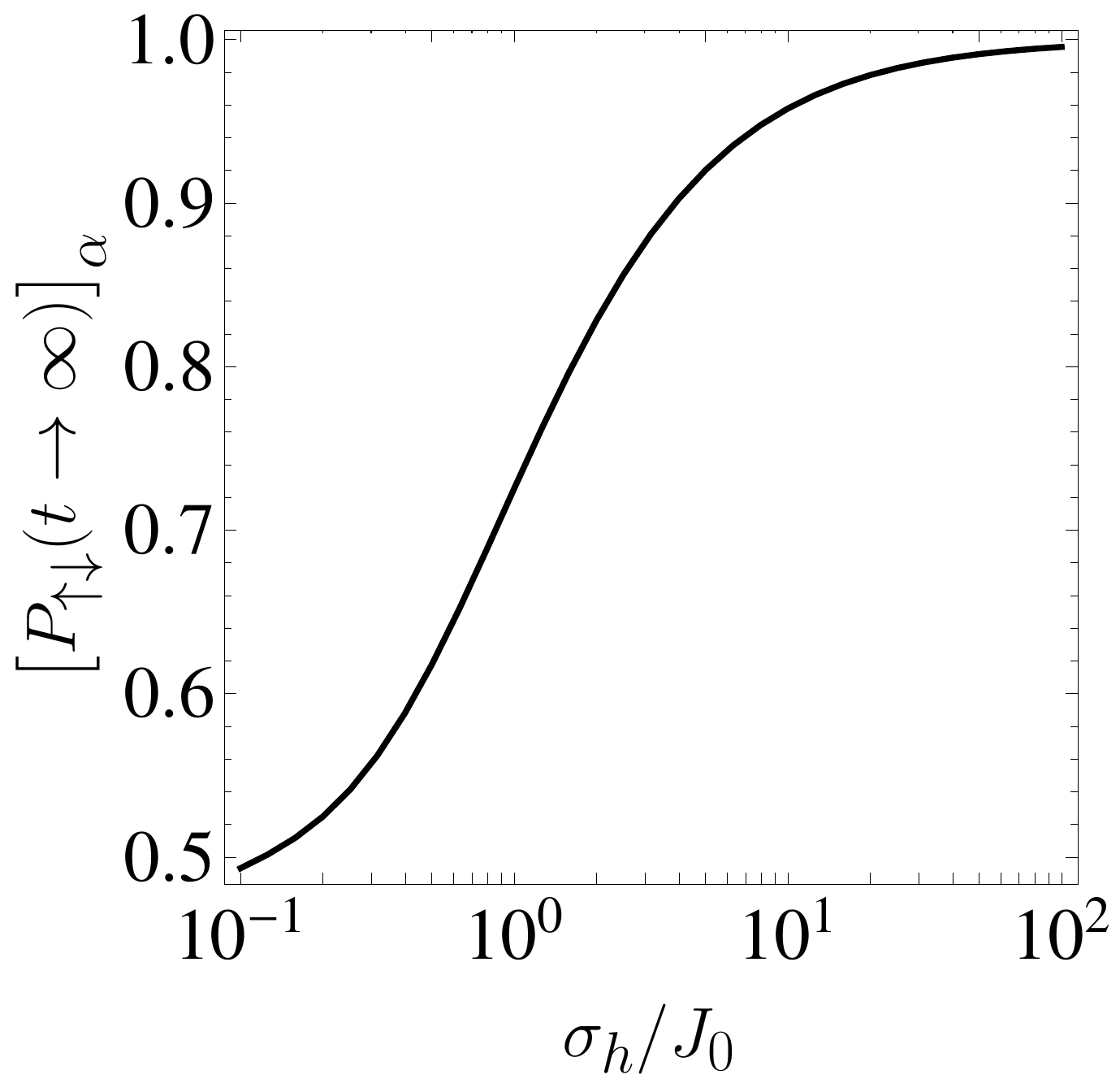}
\includegraphics[width=0.49\columnwidth]{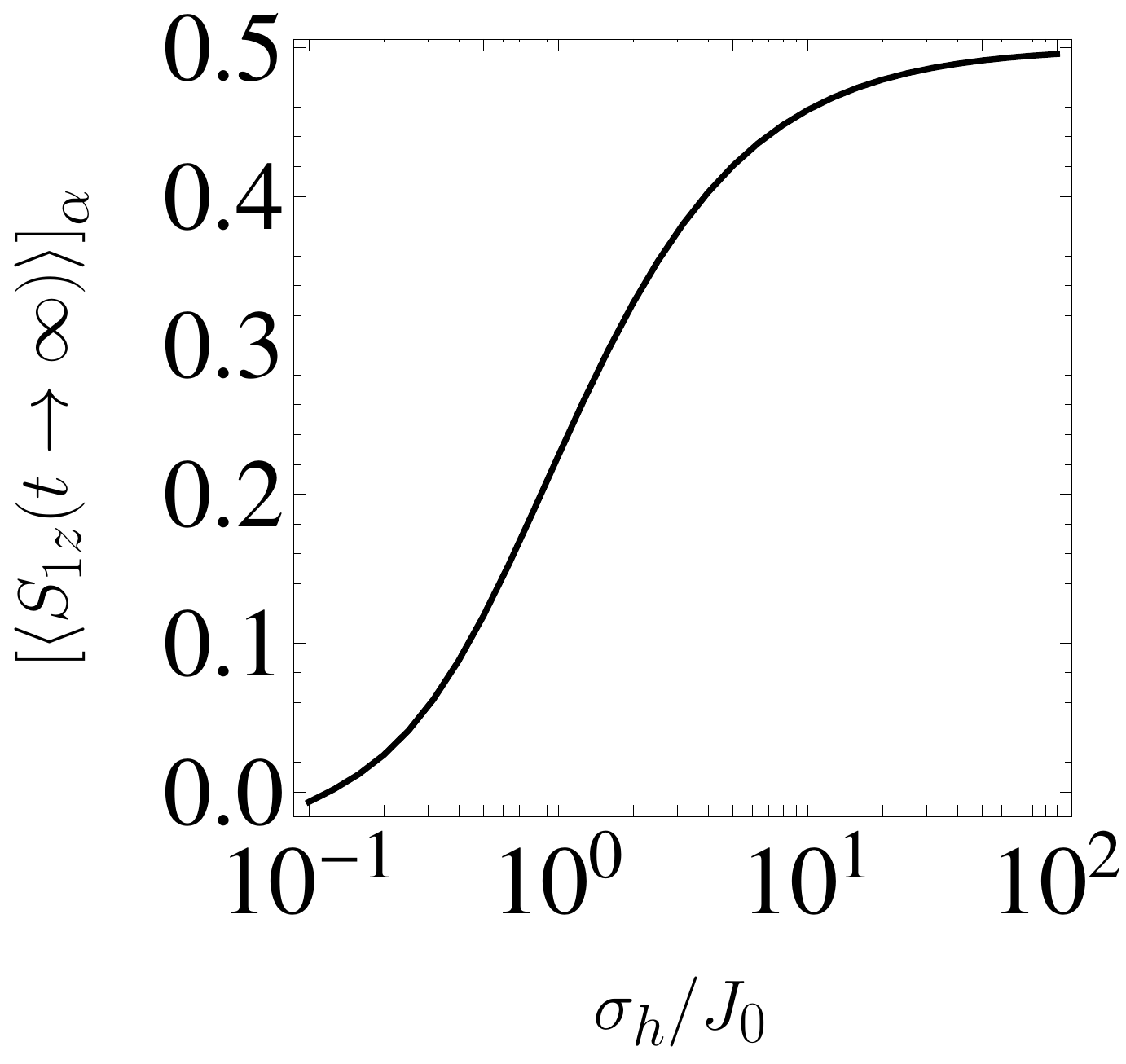}
\caption{Plot of the steady-state disorder-averaged return probability, $\left [P_{\uparrow\downarrow}(t\to\infty)\right ]_{\alpha}$
(left), and magnetization of spin $1$, $\left [\braket{S_{1z}(t\to\infty)}\right ]_{\alpha}$ (right), as functions of
$\sigma_h$ for $\sigma_J=0$ with the initial state $\ket{\psi(0)}=\ket{\uparrow\downarrow}$.}
\label{Fig:LT_noSigJ}
\end{figure}

We should emphasize, however, that, while exchange disorder has only a small effect on the steady-state return probability,
it has a significant effect on the dynamics.  To this end, we present some of our plots of the return probability and magnetization of
spin $1$ in an alternate fashion, in which we fix $\sigma_h/J_0$ and vary $\sigma_J/J_0$, in Figs.\ \ref{Fig:RPShUD} and \ref{Fig:MagShUD}.
We note that changing $\sigma_J/J_0$ has a large effect on how rapidly the oscillations in these quantities decay.
\begin{figure}[ht]
\includegraphics[width=\columnwidth]{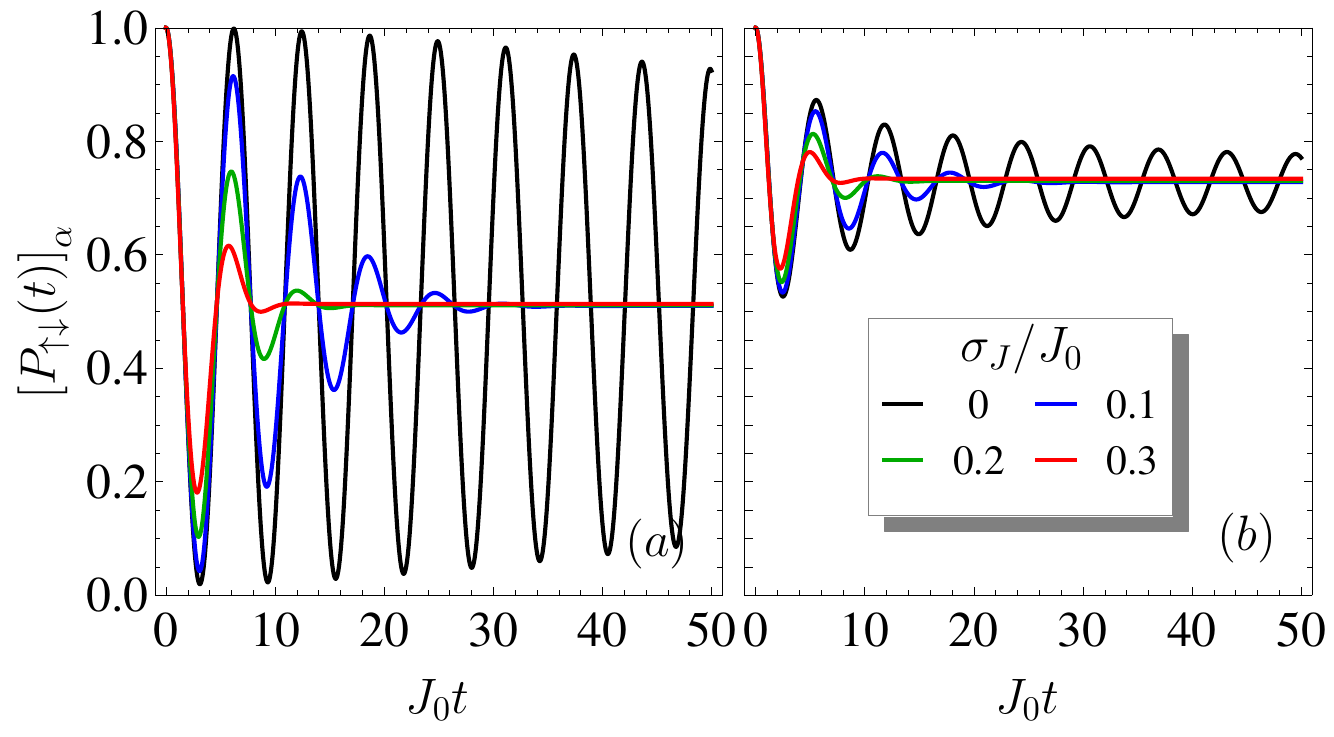}
\caption{{Plot of the disorder-averaged return probability, $\left [P_{\uparrow\downarrow}(t)\right ]_{\alpha}$, as a function
of time for different values of $\sigma_J$, as indicated in the plot, with the initial state $\ket{\psi(0)}=\ket{\uparrow\downarrow}$, for
(a) $\sigma_h=0.1J_0$, (b) $\sigma_h=J_0$.}}
\label{Fig:RPShUD}
\end{figure}
\begin{figure}[ht]
\includegraphics[width=\columnwidth]{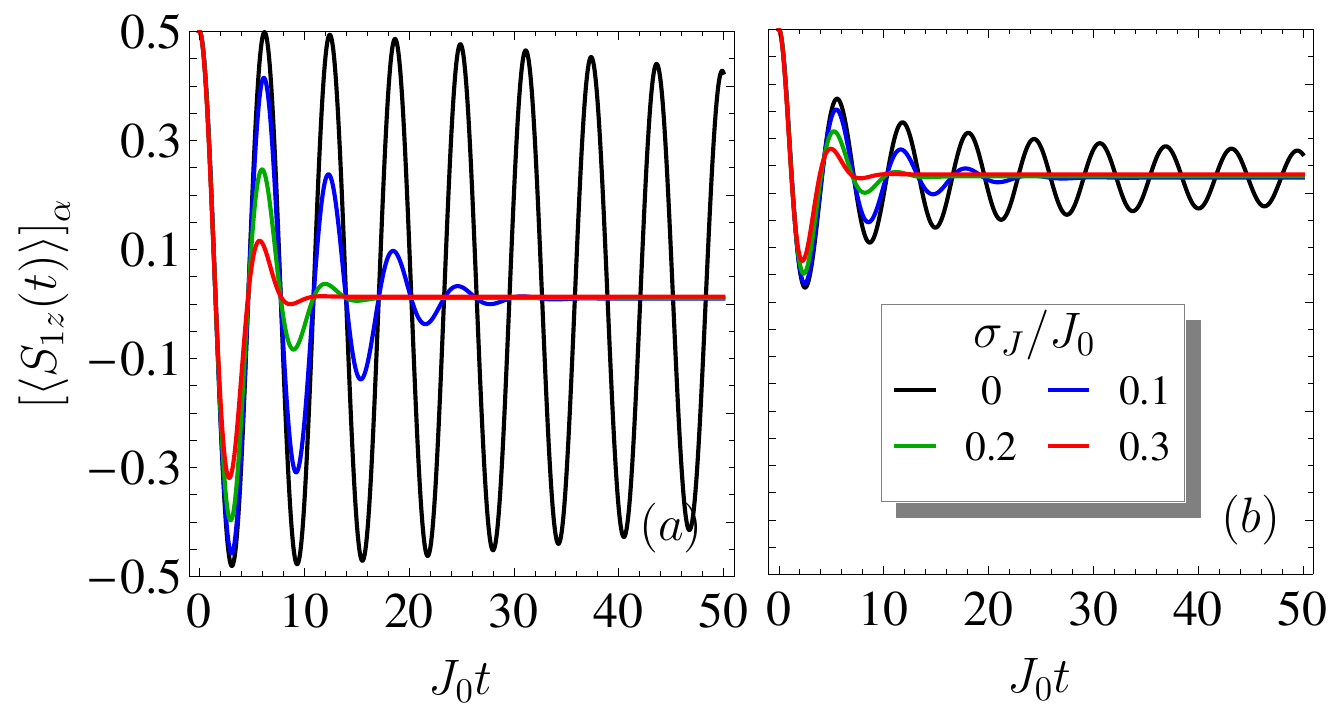}
\caption{{Plot of the disorder-averaged magnetization of spin $1$, $\left [\braket{S_{1z}(t)}\right ]_{\alpha}$, as a function
of time for different values of $\sigma_J$, as indicated in the plot, with the initial state $\ket{\psi(0)}=\ket{\uparrow\downarrow}$, for
(a) $\sigma_h=0.1J_0$, (b) $\sigma_h=J_0$.}}
\label{Fig:MagShUD}
\end{figure}

\subsection{Singlet starting state}
We now consider the same analysis done above, but taking the ``singlet'' state $\ket{\psi(0)}=\ket{S}=\frac{1}{\sqrt{2}}(\ket{\uparrow\downarrow}-\ket{\downarrow\uparrow})$
as our initial state.  In this case, the time-evolved state is given by
\begin{eqnarray}
\ket{\psi(t)}&=&\tfrac{1}{2}\left [\sqrt{1+\frac{\delta h}{\sqrt{J^2+(\delta h)^2}}}\right. \cr
&-&\left.\sqrt{1-\frac{\delta h}{\sqrt{J^2+(\delta h)^2}}}\right ]e^{iE_{+}t}\ket{+} \cr
&+&\tfrac{1}{2}\left [\sqrt{1+\frac{\delta h}{\sqrt{J^2+(\delta h)^2}}}\right. \cr
&+&\left.\sqrt{1-\frac{\delta h}{\sqrt{J^2+(\delta h)^2}}}\right ]e^{iE_{-}t}\ket{-}.
\end{eqnarray}
We find that the return probability for this state is given by
\begin{eqnarray}
P_S(t)&=&\left |\braket{\psi(t)|\psi(0)}\right |^2 \cr
&=&1-\frac{(\delta h)^2}{J^2+(\delta h)^2}\sin^2\left [\tfrac{1}{2}\sqrt{J^2+(\delta h)^2} t\right ]
\end{eqnarray}
and the magnetization of spin $1$ is given by
\begin{equation}
\braket{S_{1z}(t)}=-\frac{2J\,\delta h}{J^2+(\delta h)^2}\sin^2\left [\tfrac{1}{2}\sqrt{J^2+(\delta h)^2}\right ].
\end{equation}
As before, the magnetization of spin $2$ is given by $\braket{S_{2z}(t)}=-\braket{S_{1z}(t)}$.  We plot these expressions
for several different values of $\delta h$ in Fig.\ \ref{Fig:SingRealSing}.
\begin{figure}[ht]
\includegraphics[width=\columnwidth]{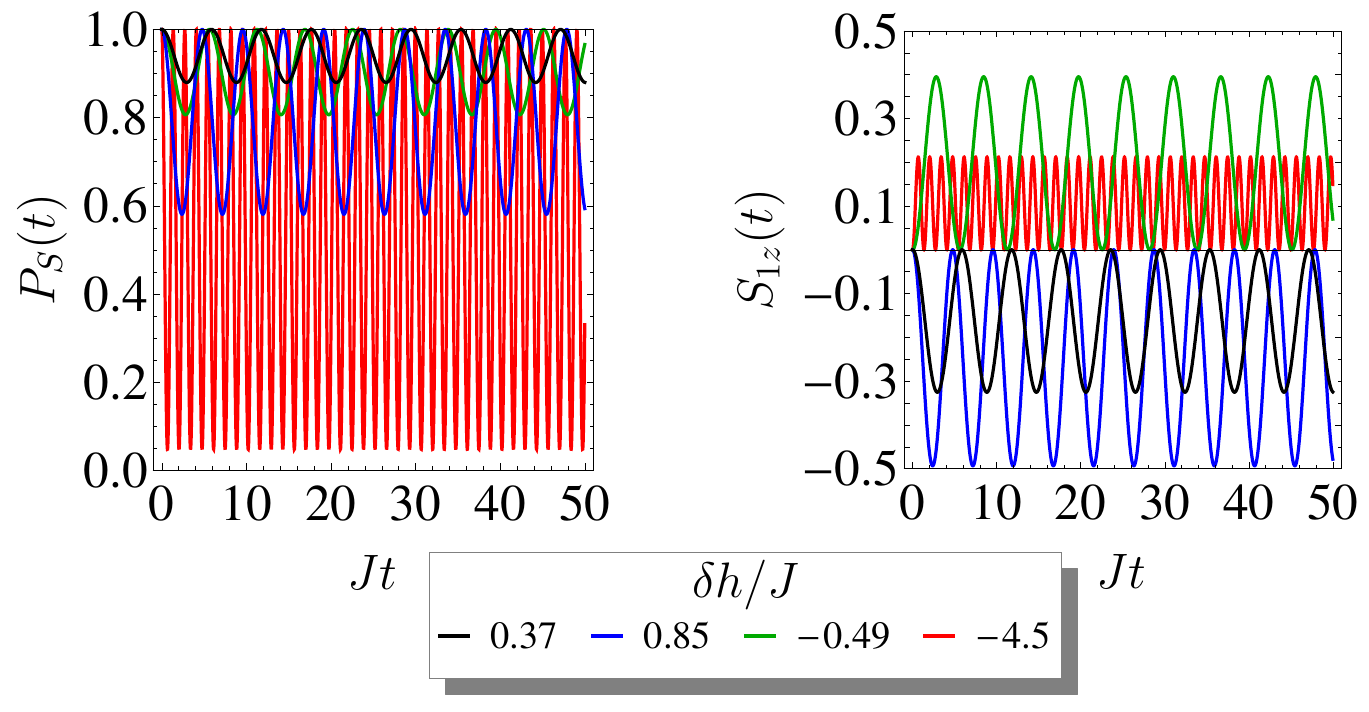}
\caption{Plot of the return probability, $P_S(t)$ (left), and of the magnetization of spin $1$,
$\braket{S_{1z}(t)}$ (right), as a function of time for several values of $\delta h$ with the initial state $\ket{\psi(0)}=\ket{S}$.}
\label{Fig:SingRealSing}
\end{figure}
The major difference between the expression for the return probability in this case and that obtained for the ``classical''
initial state $\ket{\uparrow\downarrow}$ is the presence of $(\delta h)^2$ in the numerator rather than $J^2$.  This means
that the disorder-averaged return probability will actually {\it decrease} as we increase the strength of the magnetic disorder.
This can be understood from the fact that this is an eigenstate of the Hamiltonian with the magnetic fields for each spin
set equal to each other, corresponding to zero magnetic disorder.  This is in contrast to the ``classical'' initial state
case, in which the initial state was an eigenstate of the Hamiltonian with the exchange term set to zero, corresponding to
very large magnetic disorder.  This is exactly what we find, as we will see shortly.  We also note that the magnetization
is an odd function of $\delta h$.  This, combined with Eqs.\ \eqref{Eq:DisorderAvg}--\eqref{Eq:ExchDisorderDist}, means that
the disorder average of the magnetization is zero at all times and for all values of $\sigma_h$ and $\sigma_J$; as a result,
we will be presenting disorder-averaged results only for the return probability.

We present a plot of the disorder-averaged return probability as a function of time for several values of $\sigma_h$ and
$\sigma_J$ in Fig.\ \ref{Fig:RPSJSing}.  As before, we note that the return probability oscillates and decays to a steady-state
value after a long time, and we plot these steady-state values in Fig.\ \ref{Fig:RPLT_noSigJ_ST}.  We note some major differences
between this scenario and the previous scenario, in which we use a ``classical'' initial state.  {First of all, we see
that the magnetic disorder strength seems to have a greater effect on the decay rate than in the previous case.}  We also confirm
our previous suspicion that the steady-state return probability would in fact decrease as the strength of the magnetic disorder
increases, in contrast to the ``classical'' initial state considered earlier.  {We also present plots of the return
probability as a function of time in which we hold $\sigma_J/J_0$ constant and vary $\sigma_h/J_0$ in Fig.\ \ref{Fig:RPShST}.}
\begin{figure}[ht]
\includegraphics[width=\columnwidth]{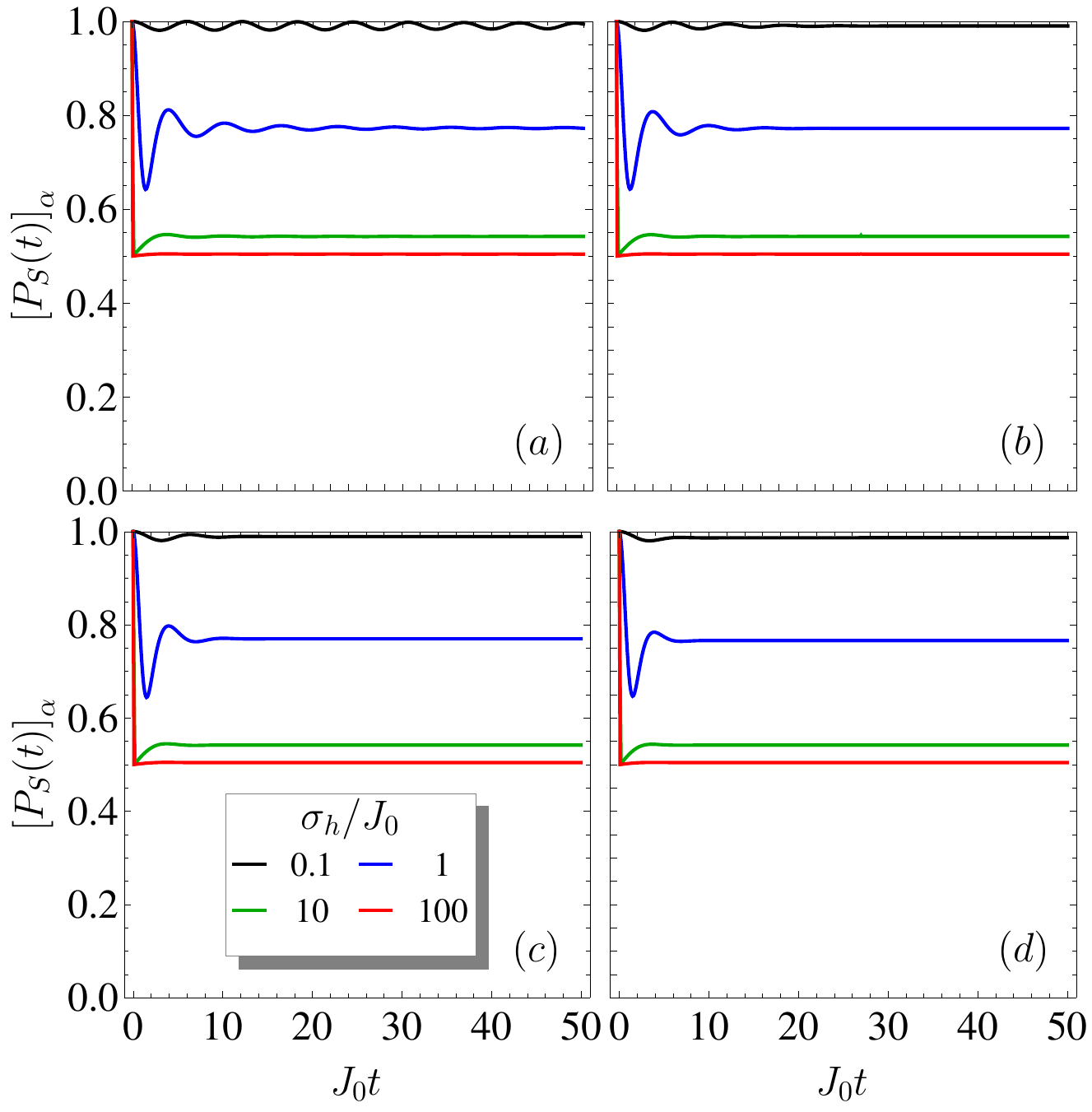}
\caption{Plot of the disorder-averaged return probability, $\left [P_S(t)\right ]_{\alpha}$,
as a function of time for different values of $\sigma_h$, as indicated in the plot, with the initial state $\ket{\psi(0)}=\ket{S}$,
for (a) $\sigma_J=0$, (b) $\sigma_J=0.1J_0$, (c) $\sigma_J=0.2J_0$, (d) $\sigma_J=0.3J_0$.}
\label{Fig:RPSJSing}
\end{figure}
\begin{figure}[ht]
\includegraphics[width=0.5\columnwidth]{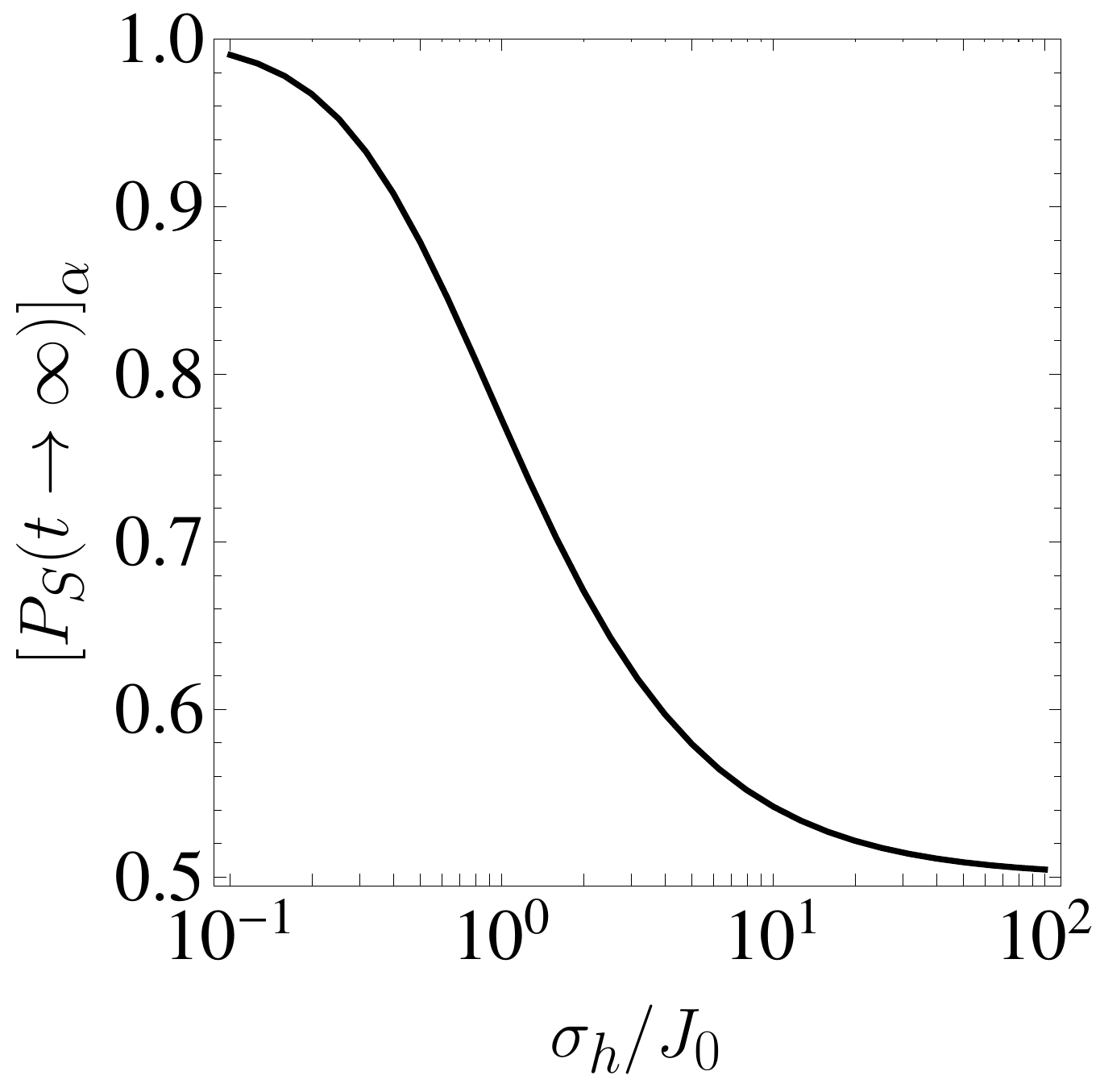}
\caption{Plot of the steady-state disorder-averaged return probability, $\left [P_S(t)\right ]_{\alpha}$
as a function of $\sigma_h$ for $\sigma_J=0$ with the initial state $\ket{\psi(0)}=\ket{S}$.}
\label{Fig:RPLT_noSigJ_ST}
\end{figure}
\begin{figure}[ht]
\includegraphics[width=\columnwidth]{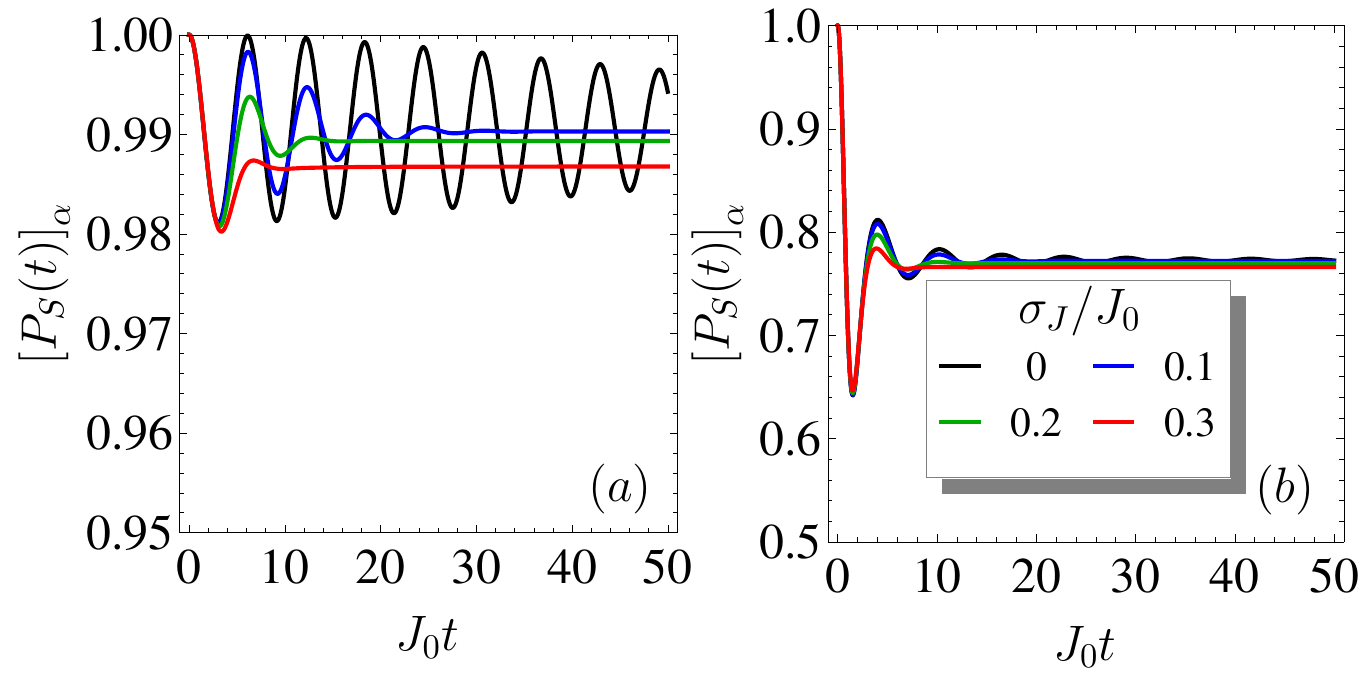}
\caption{{Plot of the disorder-averaged return probability, $\left [P_S(t)\right ]_{\alpha}$, as a function
of time for different values of $\sigma_J$, as indicated in the plot, with the initial state $\ket{\psi(0)}=\ket{S}$, for
(a) $\sigma_h=0.1J_0$, (b) $\sigma_h=J_0$.  Note the different scales on these two plots for the vertical axis compared
both to the results for the ``classical'' state and to each other.}}
\label{Fig:RPShST}
\end{figure}

\section{Ising coupling}

We now proceed to study the temporal dynamics of two capacitively coupled 
singlet-triplet qubits, described by the disordered Ising Hamiltonian in 
Eq.~\eqref{Eq:IsingH}.

We model the charge noise and the Overhauser noise using a random exchange 
coupling within each singlet-triplet qubit and a random on-site magnetic 
field, respectively.
We focus on the experimentally relevant regime dominated by the Overhauser 
disorder, and we consider only static noise, a working assumption justified 
by the slow dynamics of the nuclear spins relative to the electron spins.
The exchange couplings $J_i\ge 0$ ($i=1,2$) are drawn independently from a 
Gaussian distribution with mean $J_0$ and variance $\sigma_J^2$ truncated to 
non-negative values, whereas the magnetic fields $h_i$ ($i=1,2$) are drawn 
independently from a Gaussian distribution with mean $h_0$ and variance 
$\sigma_h^2$.
The disordered model is thus specified by the dimensionless parameters
$(\sigma_J/J_0, \varepsilon J_0, h_0/J_0, \sigma_h/J_0)$, with the first 
two controlling the inter-qubit coupling and the rest controlling single-qubit 
precession.  We note that, unlike the Heisenberg case, the uniform applied magnetic
field $h_0$ is now a relevant parameter in determining the qubit dynamics.

In the following we examine the time evolution of the return probability
$\left|\braket{\psi(t)|\psi(0)}\right|^2$ and the magnetization 
$\braket{\sigma_i^x}$ at each site $i$ for different initial states and 
different parameter settings.

\subsection{$\ket{\uparrow\downarrow}_x$ initial state}

\begin{figure}[tb]
\centering
\includegraphics[]{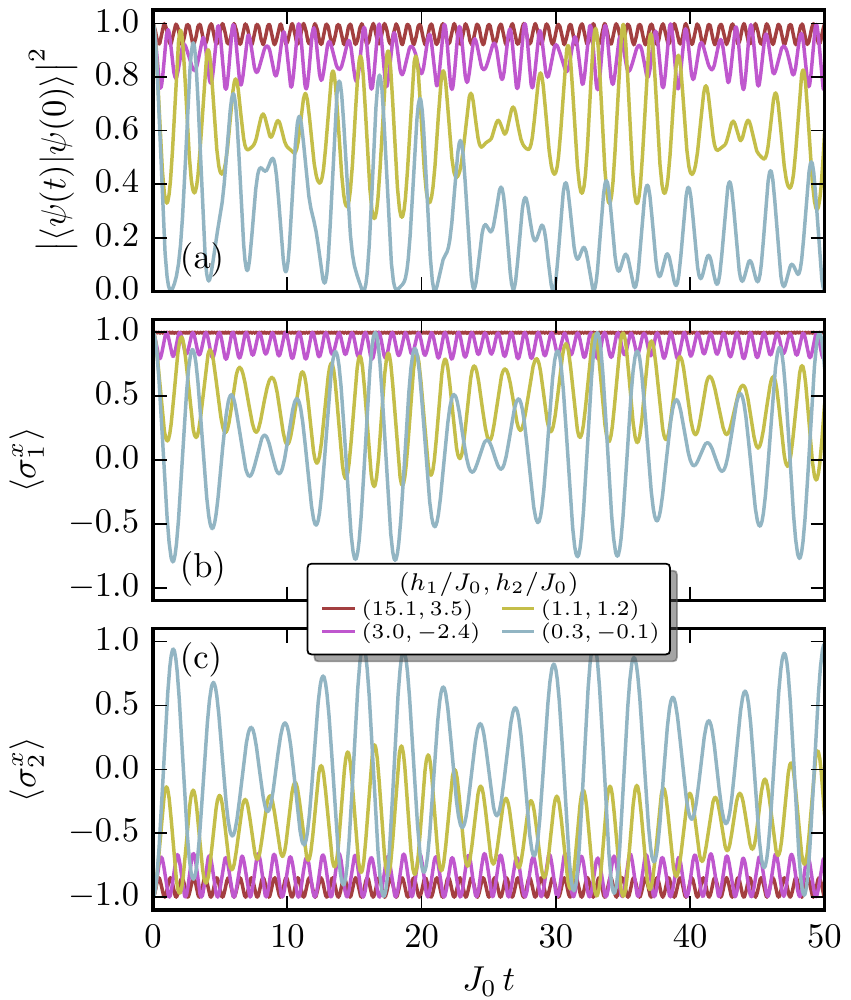}%
\caption{\label{fig:ud-single}
Dynamics of (a) the return probability and (b), (c) the single-site 
magnetizations starting from the initial state 
$\ket{\uparrow\downarrow}_x$,
for a few typical disorder realizations of the Overhauser fields $(h_1,h_2)$.
Here we use $\varepsilon=0.1J_0^{-1}$ and $J_1=J_2=J_0$, ignoring the charge 
noise $\sigma_J$.
}
\end{figure}

\begin{figure}[tb]
\centering
\includegraphics[]{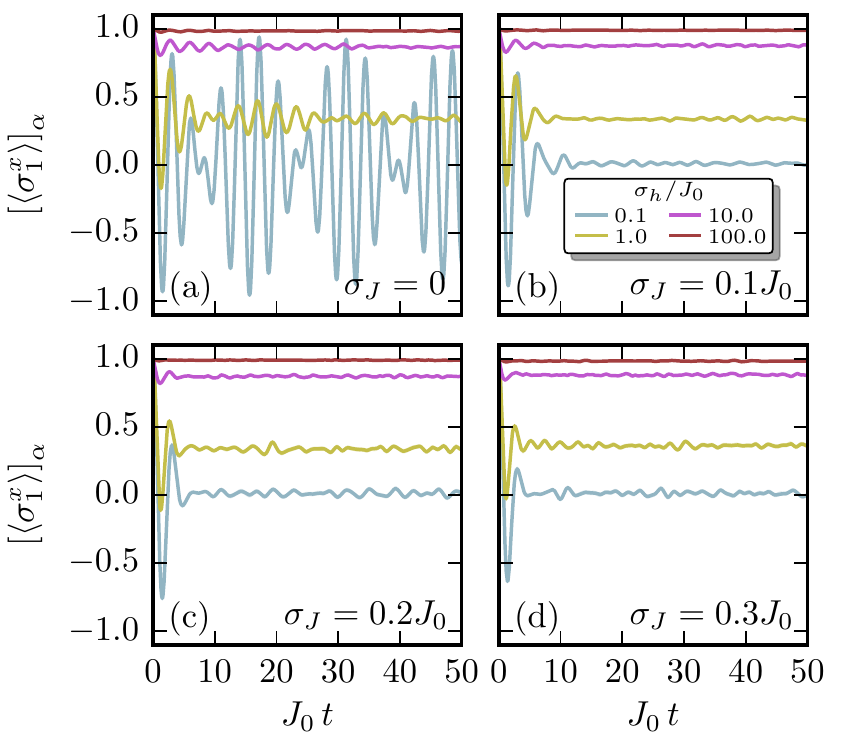}%
\caption{\label{fig:ud-mag-average-js}
Disorder-averaged dynamics of the single-site magnetization starting from the 
initial state $\ket{\uparrow\downarrow}_x$, for various strengths of the 
charge noise $\sigma_J$ and the Overhauser noise $\sigma_h$.
}
\end{figure}

\begin{figure}[tb]
\centering
\includegraphics[]{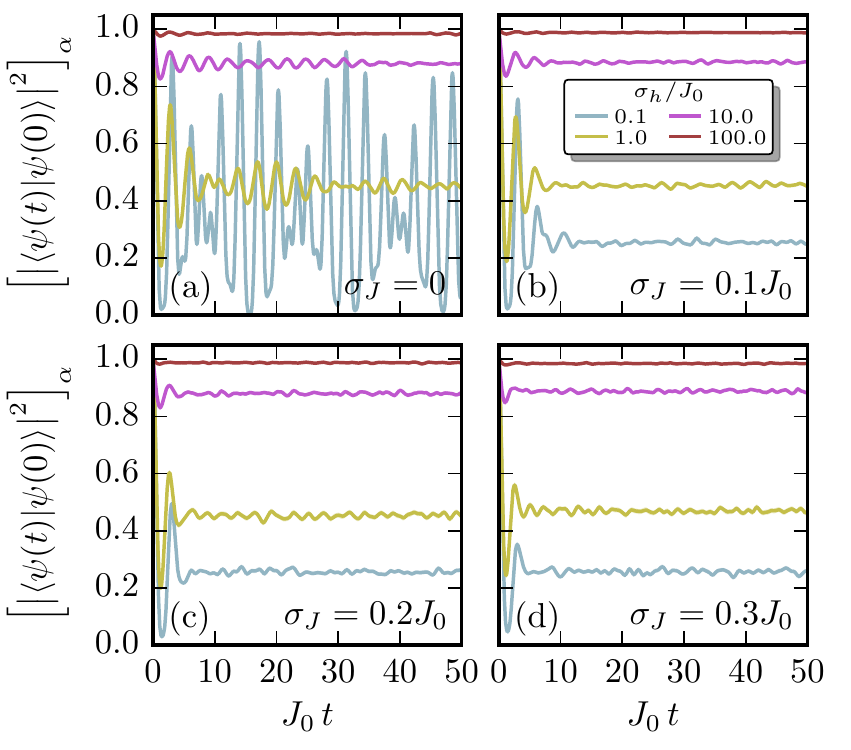}%
\caption{\label{fig:ud-rp-average-js}
Disorder-averaged dynamics of the return probability starting from the 
initial state $\ket{\uparrow\downarrow}_x$, for various strengths of the 
charge noise $\sigma_J$ and the Overhauser noise $\sigma_h$.
}
\end{figure}

\begin{figure}[tb]
\centering
\includegraphics[]{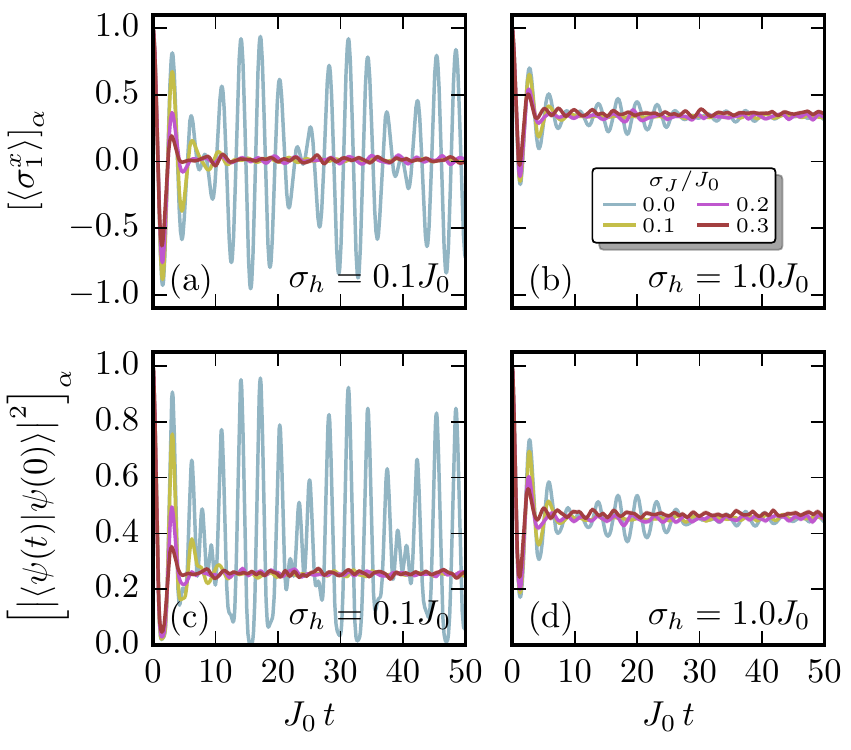}%
\caption{\label{fig:ud-js-alt}
Effect of tuning $\sigma_J$ on the
disorder-averaged dynamics of the single-site magnetization and the return 
probability, starting from the initial state $\ket{\uparrow\downarrow}_x$.
}
\end{figure}

We first consider the time evolution starting from the product initial state 
$\ket{\uparrow\downarrow}_x$.
Recall that we use $\ket{\uparrow}$ / $\ket{\downarrow}$ to denote the 
singlet/triplet state of a double quantum dot qubit.
The $\ket{\uparrow\downarrow}_x$ state we consider here is polarized in the $x$ 
direction of the singlet/triplet basis, which represents the
$\ket{\uparrow\downarrow\downarrow\uparrow}$ configuration
of the individual quantum dots of the two-qubit system.
We focus on the competition between the inter-qubit coupling and the 
Overhauser noise, ignoring the charge noise for now.
Figure~\ref{fig:ud-single} shows the coupled qubit dynamics of the return 
probability and the single-site magnetizations for a few typical 
random realizations.

When the on-site magnetic field is weak, the two-qubit dynamics are driven 
mainly by the $J_1\sigma_1^z+J_2\sigma_2^z$ exchange terms and exhibit 
fast oscillations at a frequency around $J_0$ in the absence of charge noise.
These fast oscillations are further modulated by the inter-qubit coupling term 
$\varepsilon J_1J_2\sigma_1^z\sigma_2^z$, and acquire a periodic envelope 
with a low frequency close to $\varepsilon$.
Comparing the curves in each panel of Fig.~\ref{fig:ud-single},
we find that increasing the local magnetic field $h_i$ accelerates the 
oscillation frequency and reduces the oscillation amplitude significantly.
Eventually, the Overhauser terms overwhelm the exchange coupling, and the 
coupled qubits oscillate at a high frequency around $h_i$.
At strong enough $h_i$, the magnetization at each site is essentially frozen 
to the initial value, and the return probability stays close to unity.
This comes from the fact that the $\ket{\uparrow\downarrow}_x$ 
initial state becomes an approximate eigenstate of the two-qubit Hamiltonian 
in this limit.

The temporal dynamics of the coupled qubits are characterized by persistent 
oscillations with a frequency dependent upon the Overhauser fields and the 
exchange couplings.
These fast oscillations are quickly wiped out upon averaging over the 
Overhauser disorder.
Figures~\ref{fig:ud-mag-average-js}(a) and~\ref{fig:ud-rp-average-js}(a) show 
the disorder averaged dynamics of the single-site magnetization and the return 
probability for various strengths of the Overhauser noise $\sigma_h$, in the 
absence of the charge noise $\sigma_J$.
For each parameter set we average over $10^3$ disorder realizations.
This average over the Overhauser noise introduces three changes.
First, the fast oscillations in both the return probability and the 
single-site magnetizations acquire a decaying envelope.
The gradual decay of the oscillation amplitude is clearly visible even for 
very weak disorder at $\sigma_h=0.1J_0$.
Second, the residual oscillations have a low frequency around $J_0$ controlled 
by the exchange coupling, largely independent from the Overhauser noise 
$\sigma_h$.
This can be understood as an overall destructive interference from the 
disorder average over the random magnetic field $h_i$.
Third, and most importantly, at strong Overhauser disorder $\sigma_h$, the 
local information in the initial $\ket{\uparrow\downarrow}_x$ state
persists through the time evolution, as indicated by both the return 
probability and the single-site magnetizations.

Now that we have understood the role of the Overhauser noise $\sigma_h$, we 
move on to study the charge noise $\sigma_J$.
Figures~\ref{fig:ud-mag-average-js} and~\ref{fig:ud-rp-average-js} show 
the effect of adding a moderate $\sigma_J$ on the disorder-averaged temporal 
dynamics of the return probability and the single-site magnetizations.
We find that a nonzero $\sigma_J$ further suppresses the oscillatory dynamics 
and accelerates the approach to the final steady state,
which is also clearly visible in Fig.~\ref{fig:ud-js-alt}.
This suppression comes from the mainly destructive interference between 
the random values of the exchange couplings $J_i$, and it erases the oscillations
of frequency $\sim J_0$ except for the first few periods.
This effect of $\sigma_J$ on the transient dynamics is more pronounced when 
$\sigma_h$ is weak.
Crucially, we find that a weak charge noise $\sigma_J$ does not modify the 
final steady-state values of the return probability or the single-site 
magnetizations, even when $\sigma_J$ is stronger than the Overhauser noise 
$\sigma_h$.

\begin{figure}[t]
\centering
\includegraphics[]{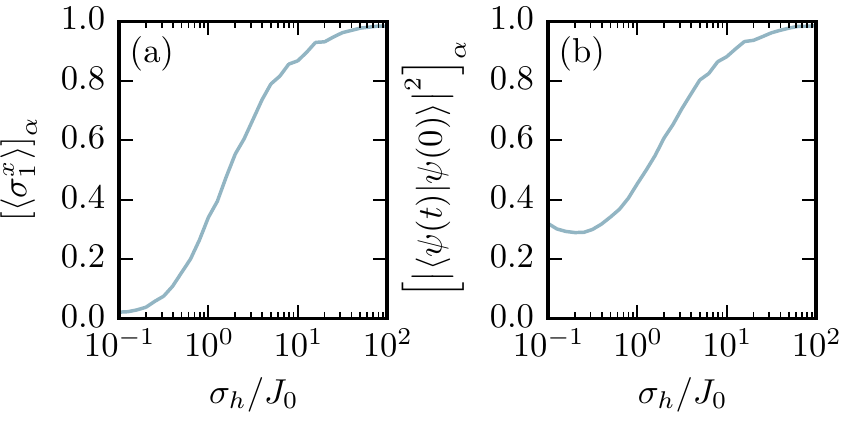}%
\caption{\label{fig:ud-mag-steady}
The Overhauser noise dependence of (a) the steady-state magnetization and (b) the 
steady-state return probability for the initial state
$\ket{\uparrow\downarrow}_x$.
}
\end{figure}

Therefore, we turn back to the Overhauser noise $\sigma_h$ and take a 
closer look at its effect on the asymptotic retention of memory of the
initial state in the final steady state.
Figure~\ref{fig:ud-mag-steady} shows the steady-state magnetization and the 
steady-state return probability, as functions of the Overhauser noise 
$\sigma_h$.
In both cases, we find a steady enhancement of the initial-state memory 
retention as $\sigma_h$ increases.
Despite the absence of a sharp transition (as is typical in such small 
systems), we note that a moderate $\sigma_h$ (around a few $J_0$) is enough to 
incur a significant boost to the preservation of the initial collective state.

\begin{figure}[tb]
\centering
\includegraphics[]{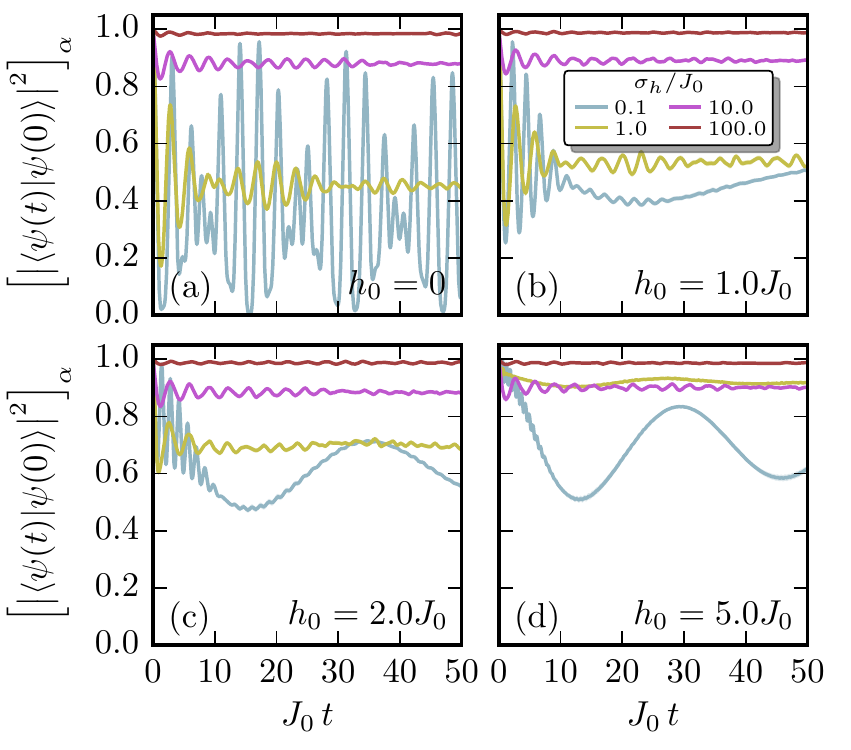}%
\caption{\label{fig:ud-rp-average-h0}
Disorder-averaged dynamics of the return probability to the 
initial state $\ket{\uparrow\downarrow}_x$ for various strengths of the 
mean magnetic field $h_0$ and the Overhauser noise $\sigma_h$, in the absence 
of the charge noise $\sigma_J$.
}
\end{figure}

Finally, we discuss the effect of adding a nonzero mean value $h_0$ to the 
random magnetic field distribution. Experimentally, this amounts to applying 
an external, uniform magnetic field to the two-qubit system.
Figure~\ref{fig:ud-rp-average-h0} shows the response of the disorder-averaged 
return probability dynamics to the mean magnetic field $h_0$.
We find that when the Overhauser noise $\sigma_h$ is relatively weak, applying 
a uniform magnetic field $h_0$ enhances the preservation of the initial 
$\ket{\uparrow\downarrow}_x$ state.
In this regime, the uniform magnetic field also suppresses the residual 
oscillations in the disorder-averaged dynamics of the return probability.

\subsection{Singlet initial state}

\begin{figure}[t]
\centering
\includegraphics[]{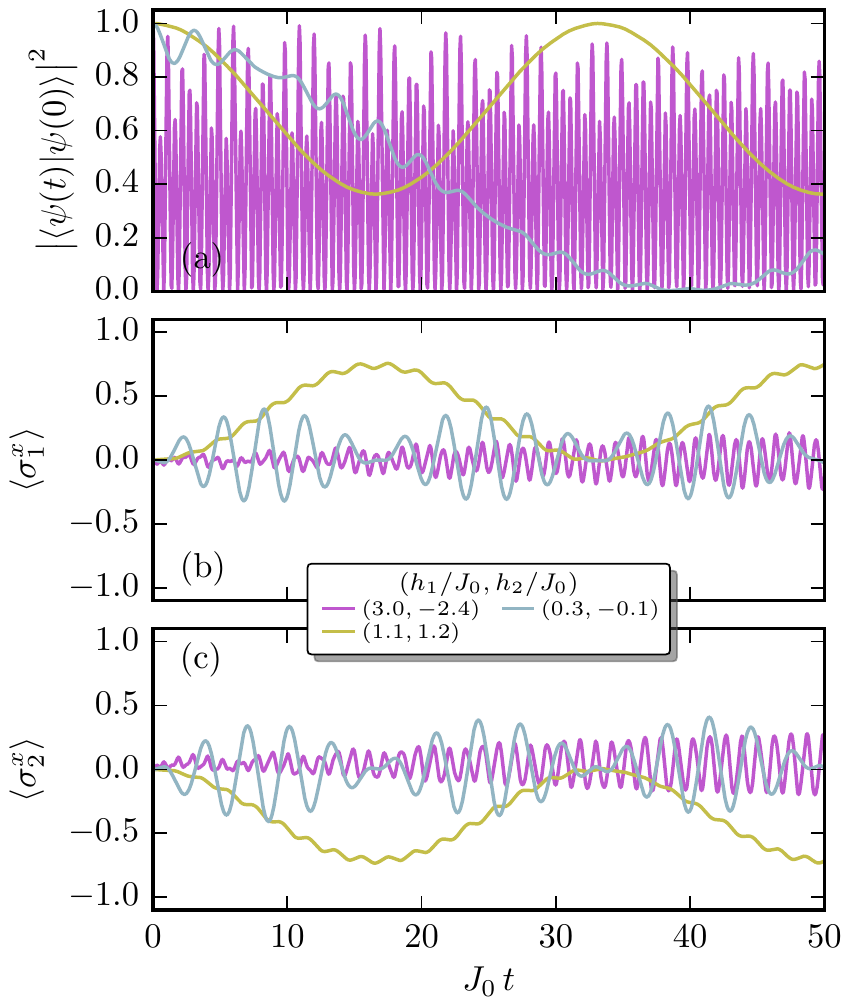}%
\caption{\label{fig:singlet-single}
Dynamics of (a) the return probability and (b,c) the single-site 
magnetizations starting from the singlet initial state $\ket{S}$
for a few typical disorder realizations of the Overhauser fields $(h_1,h_2)$.
Here we use $\varepsilon=0.1J_0^{-1}$ and $J_1=J_2=J_0$, ignoring the charge 
noise $\sigma_J$.
}
\end{figure}

\begin{figure}[t]
\centering
\includegraphics[]{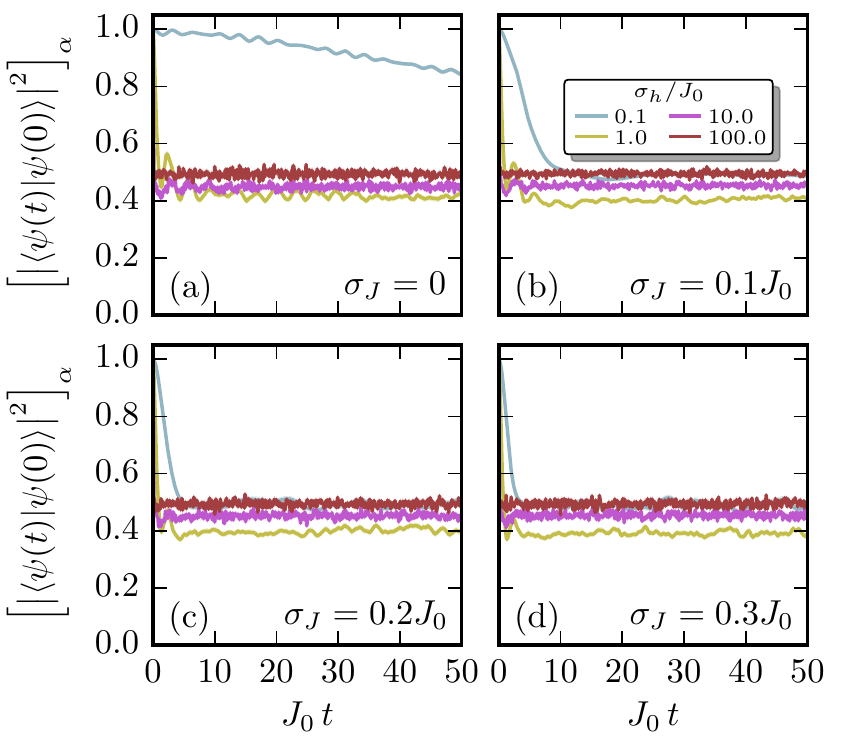}%
\caption{\label{fig:singlet-rp-average-js}
Disorder-averaged dynamics of the return probability starting from the 
singlet initial state $\ket{S}$, for various strengths of the 
charge noise $\sigma_J$ and the Overhauser noise $\sigma_h$.
}
\end{figure}

As a second example, we examine the time evolution starting from the singlet 
state
$\ket{S}=\frac{1}{\sqrt{2}}(\ket{\uparrow\downarrow}
-\ket{\downarrow\uparrow})$.
We emphasize that $\ket{S}$ is the singlet state of the pair of logical 
qubits, conceptually different from the singlet state $\ket{\uparrow}$ of each 
individual double quantum dot qubit.
In contrast to the $\ket{\uparrow\downarrow}_x$ state considered earlier, 
$\ket{S}$ is not a product state or an approximate eigenstate of the 
strongly disordered Hamiltonian.
Accordingly, the time evolution starting from $\ket{S}$ has a qualitatively 
different response to environmental noise.

Figure~\ref{fig:singlet-single} shows the dynamics of the return probability and 
the single-site magnetizations for a few typical Overhauser disorder 
realizations in the absence of the charge noise.
Notice that the initial singlet state $\ket{S}$ is an eigenstate of the 
Hamiltonian in the clean limit with a uniform exchange coupling $J_1=J_2(=J_0)$ 
and a uniform magnetic field $h_1=h_2$, with eigenvalue $-\varepsilon J_0^2$.
As the magnetic field imbalance $h_1-h_2$ increases, the return probability 
develops persistent oscillations with a large amplitude and no asymptotic 
steady state, indicating the erasure of the local information in the initial 
state.
Similar persistent oscillations are also visible in the magnetization dynamics.

This indicates that the time evolution following the singlet initial state 
$\ket{S}$ is very susceptible to disorder, in sharp contrast to the 
$\ket{\uparrow\downarrow}_x$ initial state.
To see this point more clearly, we perform an average over $10^3$ disorder 
realizations for each parameter set.
Figure~\ref{fig:singlet-rp-average-js} shows the disorder averaged dynamics of 
the return probability for various strengths of the Overhauser noise $\sigma_h$ 
and the charge noise $\sigma_J$.
(Here we focus on the return probability because the magnetization vanishes
after disorder averaging.)
We find that, for any appreciable amount of disorder, the averaged return 
probability drops to around $0.5$ within a time span as short as a few 
$J_0^{-1}$, and this descent accelerates as either type of disorder 
increases.
To understand the asymptotic behavior at long time, recall from 
Fig.~\ref{fig:singlet-single}(a) that the single-realization dynamics of the 
return probability exhibits persistent oscillations with peak-to-peak 
amplitude close to unity.
The final value of the averaged return probability around $0.5$ under strong 
disorder thus reflects the lack of an asymptotic steady state of the time 
evolution starting from the singlet initial state for each disorder realization.

To sum up, the singlet initial state shows no sign of memory retention at 
strong disorder, and the erasure of the initial state memory is accelerated by
an increase in either the Overhauser noise or the charge noise.

\section{Conclusion}
We have calculated the two-qubit dynamical evolution in the ``always on'' configuration of two coupled spin qubits in
semiconductors where the coupling could be either just an exchange coupling (i.e., Heisenberg-type) or a dipolar
capacitive coupling (i.e., Ising-type).  The central theme of the work is studying spin qubit dynamics in the presence
of both qubit interactions and qubit noise (with the noise being a random magnetic field at each qubit and/or the
interqubit coupling itself being noisy due to environmental fluctuations corresponding qualitatively to Overhauser
noise and charge noise respectively).  Our work is directly relevant for experiments on exchange-coupled spin qubits
and capacitively coupled singlet-triplet qubits when the inter-qubit coupling is kept on throughout the experiment.

Our main finding is that the disorder-averaged quantum qubit dynamics are oscillatory in time with the oscillation
decay controlled by the noise strength.  The most striking feature of the qubit dynamics in the presence of noise
that we find is a noise-induced quantum memory effect that can be directly studied experimentally using currently
existing laboratory spin qubits.  {All that one needs to be able to do is prepare either the ``classical''
$\ket{\uparrow\downarrow}$ state or the singlet state, $\ket{S}$, and be able to control the exchange coupling.
While it is true that the magnetic disorder strength, $\sigma_h$, cannot be controlled in an actual experiment, one
does have control over the dimensionless quantity, $\sigma_h/J_0$, via the exchange coupling, and this is what we
adjust in our theoretical calculations.  In experiments, one finds that $\sigma_J$ is roughly proportional to $J_0$,
so that $\sigma_J/J_0$ is a constant\cite{BarnesPrivComm}.  This would enable one to perform the necessary experiments
to test our predictions, which simply requires obtaining the return probability as a function of time, as well as the
steady-state return probability.}

In particular, we find that in many situations the steady-state probability of the final state at long times being
the initial state is very high, and that this ``memory retention'' effect, in fact, is enhanced by noise---the more
noisy the system, the more likely it is that the final two-qubit state (after the oscillations have been damped out
by noise) would essentially be exactly the same as the initial state in spite of time evolution under an arbitrary
interqubit interaction!  We provide detailed numerical and theoretical results for the steady-state return probability
(i.e., the probability that the final state of the system is the same as the initial state after the qubits have evolved
dynamically for a long time under the interqubit coupling) and the steady-state local magnetization to establish the
noise-induced memory preservation of the two-qubit system.  We emphasize that our finding of the ``memory retention'' effect
in the coupled two-qubit dynamics is quite nontrivial and counterintuitive as it happens even when the magnetic field
noise is not particularly strong, e.g., only of the order of the interqubit coupling.  We have demonstrated in a previous
work\cite{BarnesPRB2016} that ``memory retention'' effects occur in Heisenberg-coupled chains of four and six spins;
in fact, these effects are even more pronounced in the four- and six-spin systems.  These effects may be viewed as a
``remnant'' of many-body localization, which, strictly speaking, only happens in an infinitely long chain.  We have
therefore demonstrated in the current work that some qualitative manifestation of many-body localization already
happens in a two-spin system, making an experimental demonstration of such effects feasible in existing semiconductor
spin qubit platforms\cite{R2,R3,R4,R5,R6,R7,R8}.

We emphasize that the qubit dynamics depend on the initial state, and there are initial states for which memory
retention fails, but our concrete dynamical results for various initial states should help guide experimental work
studying qubit dynamics for spin qubits with exchange (``Heisenberg'') and capacitive (``Ising'') coupling.  We also
mention that our quasi-static approximation for noise is quantitatively accurate for the Overhauser noise (and is
only qualitatively valid for the charge noise affecting the interqubit coupling).  Therefore, our detailed
numerical results and the associated qualitative conclusions are valid mainly for situations where the charge noise
is not very large (or at least, not much larger than the Overhauser noise).  This makes our conclusions apply more to
GaAs-based spin qubit systems than Si systems since the Overhauser noise is typically not the limiting factor in Si-based
spin qubits.  Generalizing our work to situations involving general time-dependent noise is left for the future
since such theories would necessitate accurate quantitative information regarding the noise dynamical spectral
function not currently available in the literature.

\acknowledgements
This work is supported by LPS-MPO-CMTC.

\appendix
\section{Analytical results for the Ising model}

In this Appendix we provide an analytical formula, {previously obtained in Ref.\ \onlinecite{WangNPJQE}},
for the time evolution governed by the Ising Hamiltonian for two capacitively coupled 
singlet-triplet qubits [Eq.~\eqref{Eq:IsingH}],
\begin{equation*}
H=K\,\sigma_1^z\sigma_2^z
+J_1\sigma_1^z+J_2\sigma_2^z
+h_1\sigma_1^x+h_2\sigma_2^x,
\end{equation*}
with $K=\varepsilon J_1J_2$.

The energy eigenvalues $E_l$ ($l=1,2,3,4$) of the $4\times 4$ matrix $H$
are given by the roots of the characteristic polynomial
\begin{equation}
p(E)=E^4-2aE^2-8bE+(K^4+2cK^2+d^2),
\end{equation}
with coefficients
\begin{equation}
\begin{aligned}
a&=h_1^2+h_2^2+J_1^2+J_2^2+K^2,\\
b&=J_1J_2 K,\\
c&=h_1^2+h_2^2-J_1^2-J_2^2,\\
d&=h_1^2-h_2^2+J_1^2-J_2^2.
\end{aligned}
\end{equation}
In terms of the quartic roots $E_l$, we can express the time evolution 
operator $U(t)=e^{-iHt}$ as
\begin{multline*}
U(t)=\sum_{l=1}^4\frac{e^{-iE_lt}}{4M_l}
\big(
M_l
+u^{x}_{1l}\sigma_1^x+u^{x}_{2l}\sigma_2^x
+u^{z}_{1l}\sigma_1^z+u^{z}_{2l}\sigma_2^z\\
+u^{xz}_{l}\sigma_1^x\sigma_2^z
+u^{zx}_{l}\sigma_1^z\sigma_2^x
+u^{xx}_{l}\sigma_1^x\sigma_2^x
+u^{yy}_{l}\sigma_1^y\sigma_2^y
+u^{zz}_{l}\sigma_1^z\sigma_2^z
\big),
\end{multline*}
with coefficients
\begin{equation}
\begin{aligned}
M_l&=E_l^3-aE_l-2b,\\
u^x_{1l}&=h_1(E_l^2-d-K^2),\;
u^x_{2l}=h_2(E_l^2+d-K^2),\\
u^z_{1l}&=J_1E_l^2+2J_2KE_l+J_1(K^2-d),\\
u^z_{2l}&=J_2E_l^2+2J_1KE_l+J_2(K^2+d),\\
u^{xz}_l&=2h_1(J_2E_l+J_1K),\;
u^{zx}_l=2h_2(J_1E_l+J_2K),\\
u^{xx}_l&=2h_1h_2E_l,\;
u^{yy}_l=-2h_1h_2K,\\
u^{zz}_l&=KE_l^2+2J_1J_2E_l-K(K^2+c).
\end{aligned}
\end{equation}

\end{document}